\documentclass[dvips,12pt]{emulateapj-rtx4}
\usepackage[english]{babel}
\usepackage{amsmath}
\usepackage{amssymb}
\usepackage{url}
\usepackage[dvips]{color}

\usepackage[dvipdfm,hypertexnames=false]{hyperref} 

\newcommand{\be}{\begin{equation}}
\newcommand{\ee}{\end{equation}}
\newcommand{\beq}{\begin{eqnarray}}
\newcommand{\eeq}{\end{eqnarray}}

\begin{document}

\title{Gravitational Waves and the Maximum Spin 
Frequency of Neutron Stars}
\author{Alessandro Patruno, Brynmor Haskell, Caroline D'Angelo}
\altaffiltext{}{Astronomical Institute ``Anton Pannekoek'', University of
  Amsterdam, Postbus 94249, 1090 GE Amsterdam, the Netherlands}

\begin{abstract}
\noindent
In this Letter we re-examine the idea that gravitational waves are
required as a braking mechanism to explain the observed maximum
spin-frequency of neutron stars. We show that for millisecond X-ray
pulsars, the existence of spin equilibrium as set by the
disk/magnetosphere interaction is sufficient to explain the
observations. We show as well that no clear correlation exists between
the neutron star magnetic field $B$ and the X-ray outburst luminosity
$L_{X}$ when considering an enlarged sample size of millisecond X-ray
pulsars.
\end{abstract}
\keywords{pulsars: individual(SAX J1808.4-3658, XTE J1814-338, IGR J00291+5934, XTE J1751-305) --- gravitational waves } 
\maketitle
\section{Introduction}

Some accreting neutron stars in low mass X-ray binaries are thought to
be spun up by transfer of angular momentum from an accretion disk. The
spin-up process brings several neutron stars (NS) to spin in the
millisecond range. We refer to these systems as millisecond X-ray
pulsars (MXPs), which comprise accretion-powered millisecond pulsars
(with pulsations formed via magnetic channeled accretion), and
nuclear-powered millisecond pulsars (with burst-oscillations observed
during thermonuclear bursts). These are very old systems with
lifetimes of several billion years \citep{vanp95}.

The spin-frequency ($\nu_{s}$) of MXPs has been measured to date 
in 22 systems (\citealt{pat10}, \citealt{pap11a}) with 
 a range between $\approx 182$ Hz and $\approx
620$ Hz. The narrow range of spin frequencies appears surprising 
considering that the binary lifetime is billion years whereas the
spin-up process operates on timescales of $10^7-10^8$ yr
\citep{whi88}.

\citet{cha03} and \citet{cha08} observed that the spin-frequency
distribution of these systems (in 2003 comprising only 11 systems) was
consistent with a uniform distribution with a cutoff frequency of 730
Hz. \citet{pat10} has shown how the same cutoff is found today with a
sample size that is twice as large. The existence
of a cutoff at such a relatively low frequency is somewhat unexpected
because the Keplerian break-up frequency for a typical NS with any
realistic equation of state is well above the 730 Hz cutoff.

\citet{whi97} (henceforth WZ97) argued that the tight range 
of spin periods over two orders of magnitude in
observed luminosity, could be explained if these systems had reached
spin equilibrium.  In this picture the stellar magnetic field
truncates the disk close to or at the corotation radius (where the
stellar spin-frequency is equal to the Keplerian frequency of the
disk), and no net angular momentum is transferred onto the star.

This interpretation leads to an unexpected correlation between the NS
magnetic field at equilibrium $B$ and the outburst X-ray luminosity
$L_{X}$. Indeed in transient systems $L_{X}$ is given by the average
mass accretion rate during an outburst, which depends on the outburst
length and on the outburst recurrence time. These quantities are set
solely by the accretion disk physics~\citep{las01}, so it seems
difficult to justify a correlation with the magnetic field.

Following an earlier suggestion of \citet{pap78} and \citet{ wag84},
 \citet{bil98} proposed a different
mechanism to limit the spin-up of MXPs, introducing
a loss of angular momentum via gravitational wave (GW) emission. 
In this scenario a correlation between B and $L_X$ is much weaker and a 
sharp spin-frequency cutoff is expected due to the
strong spin-frequency dependence ($\propto\,\nu_s^5$) of the spin-down torque.  

However, recent observational results on some MXPs have shown that the
efficiency of GW induced spin-down for MXPs might be very low.
\citet{pat10}, \citet{har11} and \citet{pap11b} measured the long-term
evolution of the spin-frequency of IGR J00291+5934, revealing both
accretion torques during one outburst and a long term spin-down during
quiescence, which they interpreted as due to magneto-dipole
torques. The combined effect of spin-up in outburst and spin-down in
quiescence leads to a timescale of several billion years for the
evolution of the 599 Hz spin-frequency of this MXP
\citep{pat10}. \citet{har11} and \citet{pap11b} also showed that GW
torques are unimportant for MXPs with frequencies up to $599\rm\, Hz$.

Another recent result (\citealt{har08, har09}, \citealt{has11}),
showed that in MXP SAX J1808.4-3658 and XTE J1814-338
stringent upper limits can be set on the accretion-induced spin-up of
the NS. \citet{has11} in particular, rejected
the scenario in which a continuous train of GWs is
responsible for balancing the spin-up process. Those authors proposed
spin equilibrium as set by the disk/magnetosphere interaction as the
most likely mechanism to explain the observed behavior of the two
MXPs.

In this Letter we revisit the analysis of WZ97 and show that:
\begin{itemize}
\item the observations suggest that a large majority of MXPs can be in spin
  equilibrium without implying a correlation between the B field and
  the X-ray outburst luminosity 
\item invoking GWs to explain the cutoff in the spin
  distribution is not necessary given 
  the uncertainty in models of more realistic disk/magnetosphere
  interactions
\end{itemize}

\section{Millisecond X-ray Pulsar Sample}

We use 18 MXPs (see Table~1) which have both confirmed
spin frequencies (see~\citealt{pat10b} and references therein) and a
realistic determination of the outburst flux (see~\citealt{wat08},~\citealt{rig11a}).

We do not use any accreting NS whose $\nu_s$ is inferred only
from observations of twin-kHz QPOs. The idea that the separation
$\Delta\,\nu$ between twin-kHz QPOs is directly related with
$\nu_{s}$ was supported by the observed ratio
$\Delta\,\nu/\nu_{s}\approx 0.5-1.0$ in a few MXPs
(\citealt{vand96}, \citealt{wij03}). It has recently been shown,
however, that twin-kHz QPOs might not be a good proxy for NS spin
frequencies and could even
be completely unrelated to $\nu_{s}$ (\citealt{men07},~\citealt{yin07}).

In comparison, WZ97 used a sample of 10 accreting NS, five
of which had spin frequencies inferred only from the observation of
twin-kHz QPOs and five from burst-oscillations.

\begin{table*}
\caption{Millisecond X-ray Pulsar Sample With Spin-Frequency and Luminosity}
\centering
\scriptsize
\begin{tabular}{lllll}
\hline
\hline
Source Name & $L_{X}$ & Spin Freq. & Outb. Flux &  Distance\\
& $\left(10^{36}\rm\,erg/s\right)$ & (Hz) & $\left(10^{-8}\rm\,erg/s/cm^{2}\right)$ & (kpc)\\
\hline 
4U 1608-522 & 40 & 620& 2.0 & 4.1(0.4) \\
Aql X-1 & 32 & 550& 1.3 & 4.55(1.35) \\
SAX J1748.9-2021 & 31 & 442& 0.4 & 8.1(1.3) \\
KS 1731-260 & 30 & 524& 0.49 & 7.2(1) \\
XTE J1751-305 & 28 & 435& 0.29 & 9(3) \\
IGR J17191-2821 & 18 & 294& 0.26 & 7.5(3.5) \\
\smallskip
IGR J17511-3057 & 15 & 244& 0.2 & 8(2) \\
MXB 1659-298 & 13 & 567& 0.08 & 12(3) \\
XTE J0929-314 & 7 & 185& 0.1 & 7.8(4.2) \\
SAX J1750.8-2900 & 7 & 601& 0.12 & 6.79(0.14) \\
GRS 1741.9-2853 & 6 & 589& 0.1 & 7.2(2.8) \\
XTE J1807-294 & 6 & 191& 0.072 & 8.35(3.65) \\
\smallskip
SAX J1808.4-3658 & 5 & 401& 0.35 & 3.5(0.1) \\
IGR J00291+5934 & 5 & 599& 0.16 & 5(1) \\
XTE J1814-338 & 4 & 314& 0.069 & 6.7(2.9) \\
SWIFT J1756.9-2508 & 3 & 182& 0.04 & 8(4) \\
HETE J1900.1-2455 & 2 & 377& 0.09 & 4.7(0.6) \\
NGC6440 X-2 & 2 & 206& 0.019 & 8.1(1.3) \\
\hline
\hline
\end{tabular}
\label{obsid}
\end{table*}

\section{Disk/Magnetosphere Interaction}
\label{dmi}

The interaction between a star's magnetic field and the
surrounding accretion disk provides a natural mechanism to limit the
spin-frequency of the star. By assuming that most systems have reached
their limiting spin-frequency, WZ97 estimated the magnetic field
strength for the sources in their sample and found an unexpected
correlation between $B$ and $L_X$. 

For strong magnetic fields, the disk will be truncated at some distance
from the star (at the {\em magnetospheric radius}, $r_m$), which is
generally estimated to be \citep{pr72}:

\be 
r_m=35 \xi \dot{M}_{-10}^{-2/7} M^{-1/7}_{1.4}R_{10}^{12/7}B_8^{4/7}\rm\,km
\label{eq:rm_simp} 
\ee 
where $\dot{M}_{-10}$ is the accretion rate in units of $10^{-10}
M_\odot/\mbox{yr}$, $M_{1.4}$ is the NS mass in units of 1.4
$M_{\odot}$, $R_{10}$ its radius in units of $10$ km and $\xi$ parametrizes the
uncertainties in evaluating the torque at the edge of the accretion
disk (thought to be in the range $\xi\approx 0.3-1$, 
\citealt{psa99}).  

Inside $r_m$ gas will be channeled onto the star, spinning the star up
at a rate: 
\be 
\dot{J}=\dot{M}\sqrt{G\,M\,r_{m}}.
\label{eq:spinup}
\ee

However, if the star is spinning very fast, the disk may be truncated
outside the corotation radius $r_c$, where the Keplerian frequency is
equal to the spin-rate of the star:
\be 
r_{co}=1683\,M_{1.4}^{1/3}\nu_{s}^{-2/3}\mbox{km}\label{eq:rco},
\ee 
and the spinning magnetosphere will present a centrifugal barrier that
can inhibit accretion (the so-called ``propeller regime'', \citealt{ill75}). This suggests that the star will eventually spin at a
rate such that $r_m \simeq r_{co}$ (``spin equilibrium'').

Under this assumption, the magnetic field at equilibrium can be estimated as:
\begin{equation}
B=8.8\times\,10^{10}\,\xi^{-7/4}\dot{M}_{-10}^{1/2}M_{1.4}^{5/6}R_{10}^{-3}\nu_{s}^{-7/6}\rm\,G.\label{eq:Beq}
\end{equation}

The ten accreting NS used by WZ97 showed a spread in luminosities over
two orders of magnitude whereas the spin periods clustered between
2.8 and 3.8 s (263$-$362 Hz). Equation \ref{eq:Beq} shows that the
small range of spin periods over a large span in luminosity can be
explained within the spin equilibrium scenario only if $B\propto
L_{X}^{1/2}$ (under the reasonable assumption that
$L_{X}\propto\,\dot{M}$). However, the sample used by WZ97 suffered of
two serious biases:
\begin{itemize}
\item \textit{all} bright systems have NS spin frequencies determined
only from twin-kHz QPOs, which today is not considered a robust method
\item in two cases the spin period was taken at
  \textit{twice} the observed value to match the $\Delta\nu$ of twin-kHz QPOs.
\end{itemize}

If we remove the first bias by ignoring twin-kHz QPO sources, then the
luminosities span less than one order of magnitude instead of two.  By
removing also the second bias by using the observed spin period
(instead of twice its value), the spin periods span a range between
1.7 and 2.8 ms (362$-$589 Hz). The difference between the minimum
and maximum spin frequency therefore becomes more than twice the value
used by WZ97.
Therefore there is no clustering of MXP spin frequencies
over two orders of magnitude in luminosity as reported by WZ97. 

\begin{figure*}[]
  \begin{center}
    \rotatebox{0}{\includegraphics[width=0.9\textwidth]{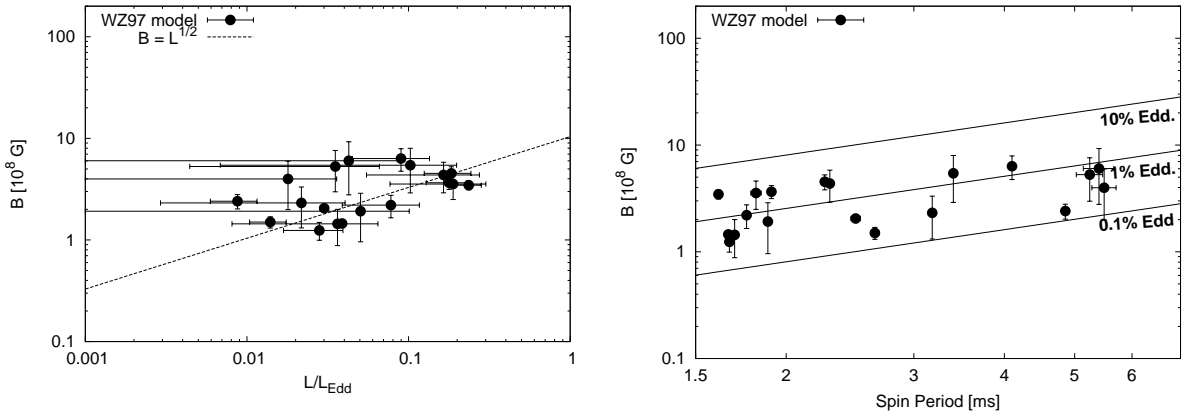}}
  \end{center}
  \caption{\textbf{Left Panel:} Inferred magnetic field at equilibrium
    vs. luminosity relation for MXPs as reported in Table 1. The black
    dots refer to magnetic fields inferred under the hypothesis of
    spin equilibrium $r_m=r_{co}$. The Eddington luminosity for a 1.4
    $M_{\odot}$ is $L_{Edd}=1.8\times10^{38}\rm\,erg\,s^{-1}$.
    \textbf{Right Panel:} Inferred magnetic field at equilibrium
    vs. spin period. The plot uses the same sources and symbols as in
    the left panel. The three lines refer to luminosities of 0.1\%,
    1\% and 10\% Eddington.}\label{fig:B-L}
\end{figure*}

We can go further and repeat the analysis of WZ97 on the enlarged
sample of MXPs given in Table 1, with the spin periods spanning now an
even broader range between 1.6 and 5.5 ms (182-620 Hz).  Assuming each source is in
equilibrium and the magnetic field is given by Eq. \ref{eq:Beq}, we see
no convincing correlation $B\propto\,L_X^{1/2}$ (see
Figure~\ref{fig:B-L}).  The field at equilibrium has now a large
spread which depends on the broad scatter in observed
$\nu_{s}$. Fig.~\ref{fig:B-L} shows also that the field is not
stronger for bright sources and weaker for faint sources as was
inferred by WZ97 (compare with Figure~1 in WZ97).

\section{Realistic models of disk/magnetosphere interactions}
\label{ref-mod}

The calculation in \S~\ref{dmi} does not incorporate the
considerable uncertainties in calculating how angular momentum can be
extracted from the star, which will strongly affect the spin
equilibrium frequency. Here we revisit a more realistic model for the
interaction between an accretion disk and the magnetic field, and
demonstrate how these uncertainties can lead to a large variation in
equilibrium spin periods for a given $B$ and $\dot{M}$. 

The estimate for the magnetospheric radius given by Eq.~\ref{eq:rm_simp} 
does not consider the relative rotation between the
star and the disk, and as such is only applicable when the disk is
truncated well inside the corotation radius (that is, the magnetic
field is rotating relatively slowly compared with the inner parts of
the disk). As the disk moves outward, the relative rotation between
the disk and star becomes critical in determining the location of
$r_m$ and the disk structure. Once $r_m$ is very close to $r_{co}$,
Eq.~\ref{eq:rm_simp} fails completely, since now the magnetic field
presents a centrifugal barrier to accretion. \cite{dan10} demonstrated
that for $r_{m} \simeq r_{co}$, the angular momentum added by the
disk-field interaction can change the disk structure and lead to a
disk truncated just outside $r_{co}$ without any accretion onto the
star or outflow.

Calculating the spin rate in equilibrium for a given accretion rate is
thus not as simple as equating Eq. \ref{eq:rm_simp} with
Eq. \ref{eq:rco}, and requires a good understanding of the spin-down
torques, which are subject to large uncertainties. For the picture
presented in \cite{dan10} the spin-down torque
 comes from the interaction between the disk and field outside
co-rotation. This leads to a spin-down torque of magnitude: 
\be
\dot{J} = \eta\frac{\Delta r}{r_{\rm m}}\frac{\mu^2}{r^3_{\rm m}},
\label{eq:spindown}
\ee where $\mu=B^{2}R^3$ is the magnetic dipole moment, $\eta \sim 1$
is a dimensionless coefficient that quantifies the strength of the
disk-field coupling, and $\Delta r/r_{m} < 0.3$ indicates the radial
extent of the coupling between the star and the disk. Both these
parameters will be dependent on the detailed disk-field coupling, and
may vary between systems as a result of, e.g., the magnetic field
inclination or the large-scale magnetic field configuration
\citep{dan11}. An additional complication in predicting the amount of
spin-down torque is the appearance of episodic cycles of accretion
\citep{spr93,dan10, dan11}, in which mass accumulates just outside
$r_c$ and is periodically dumped onto the star.  \citet{dan11} have
found that the net amount of spin-down in cyclic accretion can be up
to 100\% larger than steady accretion at the same accretion rate.

\begin{figure}[]
  \begin{center}
    \rotatebox{-0}{\includegraphics[width=1\columnwidth]{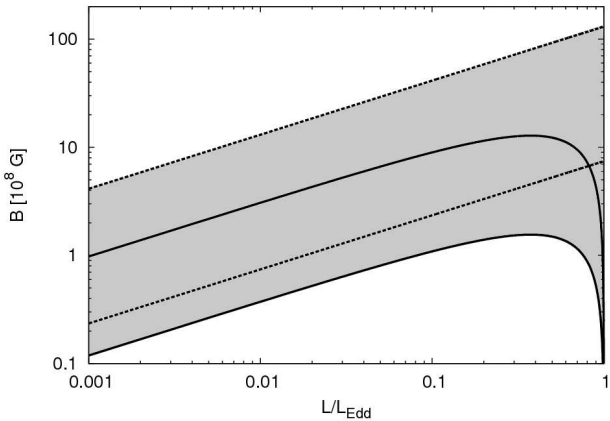}}
  \end{center}
  \caption{Allowed parameter space (shaded gray area) between the
    inferred magnetic field at equilibrium and the source luminosity
    for a hypothetical 730 Hz accreting NS. The top and bottom dashed
    lines are obtained by equating the spin-down in
    Eq.~\ref{eq:spindown} and the spin-up in Eq.~\ref{eq:spinup} and
    are calculated for $\eta\frac{\Delta\,r}{r}=10^{-3}$ and $0.3$,
    respectively. The solid lines refer to the radiation pressure
    dominated disk solution of \citet{and05}, with $\xi=0.3$ and
    1. For bright systems the inferred B field at equilibrium has an
    uncertainty that spans almost three orders of
    magnitude.}\label{fig:f2}
\end{figure}

Other sources of angular momentum loss from the star can come from
magneto-dipole radiation or stellar outflows. Magneto-dipole radiation
may dominate the spin-down for sources in quiescence \citep{har08, har09}, but
is unlikely to dominate over the disk-field interaction while the star
is in outburst. In contrast, a considerable amount of angular momentum
could be expelled via a stellar wind, depending on how much mass is
ejected in an outflow \citep{mp05}. Stellar winds have been observed
in numerical simulations to carry away a large amount of angular
momentum, and could compete or even dominate over field-disk coupling
as a source of angular momentum loss in the star\citep{rom09}.

Finally, at high accretion rates, the coupling between the disk and field can
also change the amount of angular momentum deposited onto the star
from accretion~\citep{and05}. In this case the inner parts
of the disk become radiation-pressure dominated, and the spin
equilibrium condition translates then into the relation:
\begin{equation}
B\propto \left(1-\frac{L_{X}}{L_{Edd}}\right)^{-5/6}L_{X}^{1/2}\nu_{s}^{-7/6}
\end{equation}
(see Eq. 38 and 45 in \citealt{and05}). 

Summarizing, the value of $\eta\frac{\Delta\,r}{r}$ can be different
in different systems and this might introduce a spread of B fields at
equilibrium up to an order of magnitude. For bright sources this
spread can increase up to three orders of magnitude because of the effect
of radiation-pressure dominated disks.  Therefore we would not expect
to observe $B \propto L^{1/2}_X$ even for sources that were all
spinning at the same frequency.  We show this in Fig.~\ref{fig:f2}
where we have chosen an hypothetical accreting NS spinning at 730 Hz
and a range for the parameter $\eta\frac{\Delta\,r}{r}$ in
Eq.~\ref{eq:spindown} between 0.001 and 0.3 \citep{dan11}. We have
also included the possibility that the disk becomes radiation pressure
dominated as described in \citet{and05}, with $\xi=0.3$ and 1.

The allowed region in Fig.~\ref{fig:f2} shows that a B field of
$\sim10^8$ G is already sufficient for an MXP spinning at 730 Hz at
equilibrium regardless of the source luminosity. Furthermore, since
$B\propto\nu_s^{-7/6}$, any accreting NS spinning above 730 Hz also
requires fields of the order of $10^8$ G or lower at equilibrium.
Therefore there is no need for bright systems like Sco X-1 to have
larger B fields at equilibrium than faint sources.

\section{Discussion}

In this Letter we have re-assessed the possibility that the apparent
cutoff at 730 Hz in the spin distribution of accreting NS may be due
to these systems being at or close to spin equilibrium, as set by the
disk/magnetosphere interaction. This possibility was examined in
detail by WZ97, who concluded that such a scenario would require an
unexpected correlation between the magnetic field of the star and the
outburst luminosity. Their analysis was, however, biased by the
relative scarcity of observations and by the inclusion in their sample
of bright systems for which the spin-frequency was inferred from the
separation of the twin-kHz QPOs, a method which is now known to be
unreliable (\citealt{men07}). Nevertheless, the work of WZ97 did lead
to the suggestion that the disk/magnetosphere interaction alone was
insufficient to explain the spin-frequency distribution of accreting
NS, unless the $B$ field at equilibrium was correlated with the X-ray
luminosity of the binaries. To avoid this correlation, \citet{bil98}
proposed that GWs may remove angular momentum at a rate sufficient to
explain the observed spin distribution.

Repeating the analysis of WZ97 with the extended
sample of systems available today and removing the twin-kHz QPO systems,
we have shown that
no strong correlation between the magnetic field strength and
luminosity is required to explain the observed spin distribution. This
is due to two main effects. 
First the spread in observed spin frequencies has
considerably increased since the work of WZ97, leading to a larger
scatter in the values for the magnetic field strength required for
equilibrium, even in the simplest disk/magnetosphere
picture. On the other hand the disk structure itself and the
source of spin-down torque vary from system to system, leading to a
considerable range of equilibrium B for a given $\nu_S$ and $L_x$.

From an observational point of view, there is evidence that two of the
MXPs may be close to spin equilibrium during the outburst
\citep{has11}, while the MXP IGR J00291+5934 (and to some extent XTE
J1751-395, \citealt{rig11b}) exhibits a saw-tooth like behaviour, with
the spin-up during the outburst nearly balanced
by the spin-down during quiescence (\citealt{pat10} \citealt{pap11b},
\citealt{har11}). This behaviour suggests these systems are
in long-term spin equilibrium, (\citealt{els80}, \citealt{lam05}), a
scenario which would be consistent with the inferred magnetic field of
IGR J00291+5934 ($B\approx 2\times 10^8$ G) and XTE J1751-305
($B\approx 4\times10^8 \rm\,G$). 

Furthermore, \citet{has11} have shown that GW emission
is not consistent with the timing properties and quiescent
luminosity of SAX J1808.4-3658 and XTE J1814-338. Note that
\citet{ho11} have suggested that GW emission may be required to
explain the lack of observed systems with low spin rates and long
orbital periods. This argument is, however, based on the assumption
that the spin-up torque is of the simple form in \S~\ref{dmi}
and might be necessary to reassess it with more realistic models.

The situation may be quite different for bright persistent sources
like Sco X-1 and other persistent GX and Z sources (see
\citealt{vand04} for a review on the source classification). For most
of these sources the spin-frequency of the NS is not known, and the
high accretion rate would imply spin equilibrium, for the simple
accretion torque of \S~\ref{dmi}, well above the Keplerian
breakup frequency for magnetic field strengths of the order of
$B\approx 10^8$ G. Although it is possible that these systems do
 contain a sub-millisecond pulsar, this might be difficult to
reconcile with the fact that the distribution of millisecond radio
pulsars also appears to have a cutoff at around $700$ Hz
\citep{hes06}. However, unlike
X-ray pulsars, the sensitivity of radio surveys to sub-ms pulsars
degrades with the unknown dispersion measure, which might partially
explain why radio sub-ms pulsars have never been found so far
\citep{hes07}.

A second possibility is that these systems have reached spin equilibrium at
$\sim700\rm\,Hz$ with a stronger magnetic field in bright systems,
as predicted by \citet{kon99}. If the $B$
field is confined to the NS crust a
positive correlation exists between the \textit{average mass
  accretion rate} (which, unlike the outburst accretion rate, does not
depend on the details of the disk instability model) and the
final field strength. In this scenario, the mechanism responsible for
the B field decay is Ohmic dissipation in the NS crust,
which is substantially accelerated with accretion. The
accreted material also pushes the current producing the B field deeper
down to the crust-core interface, where resistivity is low and Ohmic
dissipation becomes unimportant (see e.g., \citealt{rom90}).

An alternative scenario is that these bright systems may, in fact, be
emitting GWs. The higher temperatures of these systems and the
constant stream of accreted material make it likely that the star
could sustain a mountain in the crust, due to compositional and
heating asymmetries \citep{ush00} or that the magnetic field may be
distorted by the accretion flow and support a substantial quadrupole
\citep{mel05}. 

However, a plausible minimal scenario is that, as discussed in
\S~\ref{ref-mod}, the spin-up torque is much weaker than the
estimate in \S~\ref{dmi} at higher luminosities, while the
magnetic field strength is still in the region of $B\approx 10^8$
G. This could either be due to the inner parts of the disk being
radiation pressure dominated \citep{and05}, the disk/magnetosphere
coupling being weaker \citep{dan10} or to additional sources of
spin-down, such as mass outflows \citep{rom09}.

It would thus appear that the observed spin distribution is consistent
with the notion that all systems are either at, or close to, spin
equilibrium. This picture does not require a strong correlation
between the magnetic field strength and the luminosity and is
compatible with modest field strengths of the order $10^8$ G, with no
need to assume much higher field strengths for any systems. This field
strength is compatible with the $B$ field that is inferred from the
quiescent spin-down of three MXPs.

\bibliographystyle{apj}

\end{document}